\documentclass[preprint,12pt]{elsarticle}

\usepackage{amssymb}
\usepackage{amsmath}

\usepackage[noend]{algpseudocode}
\usepackage{algorithm}
\usepackage{footnote}
\usepackage{multirow}
\usepackage{tabulary}
\makesavenoteenv{algorithm}
\usepackage{tikz}

\usepackage{graphicx}
\usepackage{subcaption}

\journal{SEGAN}

\begin{document}

\begin{frontmatter}



\title{Coordination of Damping Controllers: A Novel Data-Informed Approach for Adaptability}


\author[inst1]{Francisco Zelaya-Arrazabal}

\affiliation[inst1]{University of Tennessee ={Department of Electrical Engineering \& Computer Science, University of Tennessee},
            city={Knoxville},
            state={TN},
            country={USA}}

\author[inst1]{Hector Pulgar-Painemal}
\author[inst1]{Jingzi Liu}
\author[inst2]{Horacio Silva-Saravia}
\author[inst1]{Fangxing Li}

\affiliation[inst2]{organization={Electric Power Group, LLC},
            city={Pasadena}, 
            state={CA},
            country={USA}}

\begin{abstract}
This paper explores the novel concept of damping controller coordination, which aims to minimize the Total Action metric by identifying an optimal switching combination (on/off) of these controllers. The metric is rooted in power system physics, capturing oscillation energy associated with all synchronous generators in the grid. While coordination has shown promising results, it has relied on computing linear sensitivities based on the grid model. This paper proposes a data-informed framework to accurately estimate total action and subsequently determine an optimal switching combination. The estimation is provided by a multivariate function approximator that captures the nonlinear relationship between system-wide area measurements, the status of damping controllers, and the conditions of the disturbance. 
By enabling real-time coordination, electromechanical oscillations are reduced, enhancing power system stability. The concept is tested in the Western North America Power System (wNAPS) and compared with the model-based approach for coordination. The proposed coordination outperforms the model-based approach, demonstrating effective adaptability and performance in handling multi-mode events. Additionally, the results show significant reductions in low-frequency electromechanical oscillations even under various operating conditions, fault locations, and time delay considerations.
\end{abstract}


\end{frontmatter}


\section{Introduction}
\label{Intro}
The sustained increase in renewable generation over the past decade has introduced greater variability and uncertainty in grid operation. Consequently, stability has been compromised, resulting in the sub-optimal performance of controllers.  For instance, in relation to low-frequency oscillations (LFO), damping controllers (DC) are commonly designed to damp out a targeted dominant mode, assuming a typical operating condition---required for linearization \cite{Zhang_Bose,aboul1996damping}. As a result, the DC performance is dependent on how close the actual real-time operating point is to the one assumed in the DC design stage \cite{wang2014multi}. In addition, the targeted mode must be excited by a disturbance. If an actual disturbance fails to excite the targeted mode, the DC might not effectively restrain oscillations. In fact, under certain circumstances, it might even compromise system stability. Therefore, addressing these challenges requires enabling damping control architectures—local and wide-area damping control (WADC)—with adaptability to varying operating conditions and disturbances, ultimately enhancing overall system stability.

Numerous adaptable approaches have been proposed \cite{ranjbar2021dynamic, alinezhad2020adaptive, wang2022adaptive, LiuY, ZHANG201645,bai2016design}, with two main categories: model-based and data-driven methods. Despite their promising results, each faces unique challenges. Model-based techniques rely on accurate and up-to-date data to effectively parameterize controllers, but implementing them is hampered by legacy equipment and proprietary models used in actuators \cite{chakraborty2022review}. On the other hand, data-driven approaches, which focus on real-time adjustments of controller parameters, may encounter restrictions from utility companies due to security concerns. These concerns arise from potential forced oscillations and resonance with natural modes caused by the uncertainty in the variable control tuning parameters \cite{silverstein2015diagnosing}. Additionally, as emphasized in \cite{chakrabortty2021wide}, there is a significant demand for novel adaptable solutions with more distributed resources, and minimal computational complexity and execution time. These challenges are evident in the limited presence of operational WADC solutions within current power grids \cite{uhlen2012wide, lu2012implementations}. For example, in the United States, the Pacific DC Intertie (PDCI) is the sole functioning system within the entire US grid. Operating within a high-voltage DC (HDVC) transmission line, the PDCI utilizes a filter-based approach, with its primary objective being to dampen the North-South B mode of the Western Interconnection. As detailed in previous research \cite{trudnowski2013pdci, pierre2019design}, one of the principal challenges was accurately tuning compensation for the NS-B mode, underscoring the need for involvement from multiple entities in the design, validation, and deployment phases \cite{trudnowski2017initial}. Furthermore, the complexity of the implementation process increases when considering multiple modes. This highlights the fact that approaches involving the adjustment of several tuning parameters in real-time may be subject to various restrictions imposed by independent system operators and utilities to prevent adverse effects on system stability.

Expanding on the quest for adaptable solutions, an innovative model-based approach proposed by the authors offers an alternative for adaptively optimizing the damping actions in the grid\cite{silva2020adaptive}. This approach, the coordination of DCs, efficiently manages numerous decentralized local or wide-area DCs across the grid without the need for real-time adjustment of tuning parameters. Notably, the coordination focuses on Inverter-based resources (IBR), allowing these resources to act as DCs only when beneficial for grid damping improvement. 
By not requiring changes to controller parameters, this approach proves to be non-intrusive and effective in reducing grid oscillations.

Given the complexity of the coordination problem, the model-based approach relies on sensitivities to determine optimal controller status \cite{silva2020adaptive}. However, this necessitates either pre-computation or real-time calculation of a linearized model and sensitivities, increasing complexity and potentially impacting execution time. In response, this paper proposes a data-informed coordination (DIC) approach that directly estimates the optimal on/off status of DCs, without relying on a linearized model. Key wide-area measured variables are utilized within a hierarchical control structure to estimate the Total Action (TA), capturing all system nonlinearities, specific disturbances, and operating conditions. Notably, no proprietary models are necessary, and IBR resources are utilized as DCs only when required. This work presents four main contributions: a) Adaptive Data-Informed Coordination: This approach provides real-time switching on/off conditions based on wide-area measurements; b) Coordination Structure Based on Total Action Function Approximation: A structure comprising three main parts aims to gather accurate data and commission IBR-DCs; c) Overcoming Model-Based Coordination Drawbacks: This coordination structure eliminates dependency on the linearized model and its corresponding sensitivities; and d) Improving System Damping Against Multi-Mode Events: The coordination structure enhances system damping even when various modes have been excited.

The paper is structured as follows. Section \ref{ADCDC} discusses the concepts of oscillation energy, total action, and model-based coordination (MBC). Section \ref{DDADC} presents the use of deep learning theory for data-informed coordination (DIC), while Section \ref{DDCimpl} describes its implementation. The outcomes of data-informed coordination, its time performance, and time delay evaluation are presented in Section \ref{case_study}. Finally, a discussion of the results and real deployment, along with conclusions, can be found in Sections \ref{disc} and \ref{conclusion}.
\section{Background on controllers coordination}
\label{ADCDC}
The emerging concept of controllers coordination for real-time control is possible due to the use of a single metric called total action, which is derived from the system oscillation energy. Assume a system with $p$ synchronous generators (SGs). Then, the oscillation energy (E) is given by,
\begin{equation}
	\label{KinecticEnergy}
	E(t) = \sum^{p}_{j=1} H_{j} \omega_{s} \Delta \omega^{2}_{j}(t) 
\end{equation}
where the subscript $j$ represents the $j$-th SG, $w_{s} = 120 \pi$ is the synchronous speed in rad/s, $H_{j}$ is the inertia constant in s, and $\Delta w_{j} = w_{j} - w_{COI}$ the speed deviation in p.u.---the speed of the center of inertia (COI) is used, which is defined as $\omega_{COI}(t) = \sum^{p}_{j=1} H_{j} \omega_{j}(t) / \sum^{p}_{j=1} H_{j}$. The multivariate function $E(t)$ can exhibit an oscillatory pattern in time if some oscillatory modes are excited, and it converges to zero when the system has asymptotic stability. The time integral of $E(t)$ over an horizon $\tau$ is known as the action $S(\tau) = \int_{t_0}^{t_0+\tau} E(t) \,dt$. The time $t_0$ can be the time of disturbance, or any other after this such as the time when the first measured data is available. In a stable system, the TA is defined as,
\begin{equation}
	\label{TotalAction}
	S_{\infty} = \lim_{\tau \to \infty}	S(\tau)  = \lim_{\tau \to \infty} \int_{t_0}^{t_0+\tau} E(t) \,dt
\end{equation}
The TA considers the entire action of the system after a disturbance, a large value of this reflects large and sustained oscillations \cite{silva2018oscillation,silva2020adaptive}. In other terms, an increased system damping would imply a low TA. This physical concept not only allows avoiding the use of arbitrary objective functions in regards to oscillation damping, but also serves as a mechanism to weight the importance of relevant oscillation modes without having to target in advance the most critical ones. By having a notion of the TA for any given disturbance, the on/off status of the damping controllers can be determined as those that lead to a minimum TA. A model-based approach has been proposed to solve the coordination problem \cite{silva2020adaptive}, which relies on a linearized model. A concise description of this approach is presented next.

If $A_{q}$ is the system matrix of the linearized model, a similarity transformation is used to convert the TA into an explicit expression in terms of the system eigenvalues and initial condition after a disturbance. With $\Lambda_{q} = M^{-1}_{q} A_{q}M_{q} = diag\{\lambda_{qi} \}$, where $\lambda_{qi}$ is the $i$-th eigenvalue and $M_{q}$ is the matrix of right eigenvectors, the TA can be estimated by,
\begin{equation}
	\label{TotalAction}
	S_{\infty} \approx -\frac{1}{2} \sum^{n}_{j=1} \sum^{n}_{i=1} \frac{z_{0i}z_{0j}g_{ij}}{\lambda_{qi}+\lambda_{qj}} 
\end{equation}
Here, $g_{ij}$ is the (i,j)-term of the transformed inertia matrix, and $z_{0i}$ and $z_{0j}$ are the i-th and j-th terms of $z_0=M_q^{-1}x_0$, with $x_0$ being the vector of state variables at time $t_0$---initial condition of the transformed and original state variables. The subscript $q$ indicates the dependency on the on/off switching status of the controllers.

\subsection{Model-based coordination based on TA sensitivities}
Assume there are $m$ DCs, and define $q_{k} \in \{0,1\} \ \forall \ k \in \{1,2,...,m\}$ as the on/off switching status of the k-th DC, i.e., $q_k$=1 if the m-th DC is activated, and $q_k=0$ otherwise. The coordination aims at increasing the system damping by specifying $q_{k} \ \forall \ k \in \{1,2,...,m\}$ based on the current operating condition and disturbance. The optimal on/off status of the DCs, $q_k^*$, is obtained by solving the binary integer problem that minimizes $S_{\infty}$. Due to the problem complexity, a sub-optimal solution can be achieved through linear sensitivities of the TA. By relaxing the switching variables $q_{k}$ and varying them in the range $[0,1] \in \mathbb{R}$, a first order Taylor expansion of $S_\infty$ around an initial switching condition $q_{k0}$ is obtained:
\begin{equation}
	\label{TotalAc_aprox}
	\!
	\begin{aligned}
		\Delta S_{\infty} & \approx \frac{\partial S_{\infty}}{\partial \hat{q_{1}}}\Delta \hat{q_{1}} + \frac{\partial S_{\infty}}{\partial \hat{q_{2}}}\Delta \hat{q_{2}} + ... + \frac{\partial S_{\infty}}{\partial \hat{q_{m}}}\Delta \hat{q_{m}}
	\end{aligned}
\end{equation}
The partial derivatives are called total action sensitivities (TAS) and are explicitly estimated using the linearized model and its  eigenvectors, eigenvalues and derivatives \cite{silva2018oscillation}.

\begin{figure*}[b]
	\centering
    \includegraphics[width = 0.75\textwidth]{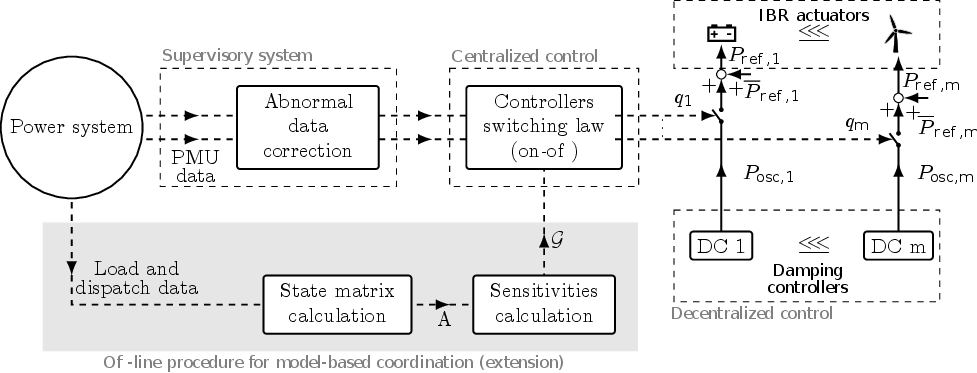}
	\caption{DCs coordination framework}
	\label{DB_ADDC}
\end{figure*}

For a given operating condition and disturbance, while a gain increment of the k-th DC would worsen the dynamic performance if its TAS is positive, an enhanced performance would be achieved if its TAS is negative. The sub-optimal solution is obtained by applying the following on/off switching logic for all $k \in \{1,2,...,m\}$: $\mathbf{Switch \ on} \ \{ q_{k}: 0 \rightarrow 1 \Longleftrightarrow \frac{\partial S_{\infty}}{\partial \hat{q_{k}}} |_{\hat{q_{k}}} < 0 \}$ and $\mathbf{Switch \ off} \ \{ q_{k}: 1 \rightarrow 0 \Longleftrightarrow \frac{\partial S_{\infty}}{\partial \hat{q_{k}}} |_{\hat{q_{k}} } > 0 \}$. The implementation of this control logic requires offline and online stages. For each predefined operating condition and for each predefined disturbance, scenario $k$, the offline stage calculates the TAS for all DCs and stores them as a single data collection $\mathcal{G}_k$ contained in the superset $\mathcal{G}$. The online stage is implemented as a two-level hierarchical control: The first level is the decentralized control, which corresponds to the traditional damping control, and the second level is the centralized control, which corresponds to the DCs coordination through the TAS. For a real operating condition and disturbance, a close enough scenario is picked from the superset $\mathcal{G}$. If nothing close is found, a fast estimation of the TAS would be required. Test data showed that even a 2-second time window for detection, calculation, and DCs switching action would lead to a diminished but still improved dynamic response. Despite the approach's adaptability to disturbances and operating conditions, its implementation can be cumbersome due to the need of large data storage and fast computational capabilities. Without relying on a system model, the proposed data-informed coordination in this paper not only can ease the real-time implementation with a direct estimation of the optimal on/off status of DCs, but can also improve effectiveness by a prompt control action, and accuracy by capturing all system nonlinearities. Fig. \ref{DB_ADDC} shows the control framework, and indicates the extension needed for the model-based coordination. The data-informed coordination does not require this extension and can be implemented directly as a new controller switching law.
\section{Data-informed DC's Coordination}
\label{DDADC}
The motivation for using the TA as a performance index is to have a unique function that captures the spectrum for electro-mechanical dynamics and reflects the impact of control actions. The model-based approach, presented above, serves as a proof of concept of the coordination problem using a TA linear estimation. The model-based coordination is only accurate in the vicinity of the predefined conditions. Moreover, for higher effectiveness, more scenarios for operating conditions and disturbances would be required, which can significantly increase the cardinality of the superset $\mathcal{G}$. Therefore, a data-informed approach is an ideal solution to these issues. In that sense, a nonlinear function approximator powered by a Deep Neural Network (DNN) can be used to provide the complex nonlinear estimation involving the TA action dynamic. Finally, the optimal on/off status of the DCs can be determined by seeking the minimum of the TA function.

\subsection{Total action function approximator}
\label{DNNasTAaprox}
Based on the universal approximation theorem \cite{nielsen_1970, zhao2021deep}, despite nonlinearity and intrinsic complexities, there must be a DNN capable of capturing the relationship between the TA and key features. This is a regression problem, where the TA estimation $\hat{S}_{\infty} = f(\mathbf{\mathbf{r}})$ will be given by the DNN function approximator, $f(x)$, based on the vector of features, $\mathbf{r}$. The approximator is a nested function, where each one of its layers is formed by artificial neurons. Assume $l_n$ layers with vector functions $\mathbf{f}_{1}$, $\mathbf{f}_{2}$ ... $\mathbf{f}_{l_n}$, 
the function approximator is given by \cite{burkov2019hundred},
\begin{equation}
	\label{F_nest0}
	\hat{S}_{\infty} = \mathbf{f}_{l_n} \circ \hdots \circ \mathbf{f}_{2} \circ \mathbf{f}_{1}(\mathbf{r} )
\end{equation}
Each of the vector functions has the form:
\begin{equation}
	\label{F_nest}
	\mathbf{f}_{l} = \mathbf{g}_{l}(\mathbf{W}_{l}\mathbf{r}+\mathbf{b}_{l})
\end{equation}
where $l$ is the layer index, $\mathbf{W}_{l}$ is the matrix of weights, $\mathbf{b}_{l}$  is the vector of biases, and $\mathbf{g}_{l}$ is the activation function---in charge of providing the nonlinear behavior of the DNN. Note that $\mathbf{f}_{l}$ is a function vector. Each one of its entries is an artificial neuron defined as $a_{l,u}= g_{l}(\mathbf{w}_{l,u}\mathbf{r}+b_{l,u})$, where $u$ is the index of the neuron and $\mathbf{w}_{l,u}$ is the corresponding row vector from $\mathbf{W}_{l}$. In the particular case of the output layer, there is just one neuron, $u=1$. Thus, the function provides a scalar value.

For an accurate estimation, the DNN must be trained to approximate the values from the training data set through the calculation of the optimal parameters for each layer, i.e., $\mathbf{W}_l, \ \mathbf{b}_l \ \forall \ l$. Let $L_{DNN}$ be the loss function defined as,
\begin{equation}
	\label{loss}
	L_{_{DNN}} = \frac{1}{N} \sum_{i=1}^{N} ((\hat{S}_{\infty})_{i}-(S_{\infty})_{i})^2 
\end{equation}
where $N$ is the number of samples in the training set. Note that the loss function is the mean square error of the difference between the estimated and actual TA. To minimize the loss function and to achieve an accurate estimation, stochastic gradient descent with backpropagation can be used to find the best DNN hyperparameters. 

The features, $r$, can have an operational or control nature. To be precise, the vector $r$ is decomposed as $r=[y_0, \gamma_k]^T$. The vector $y_0$ has features relevant to the dynamic response such as the speed deviation from each SG. The subscript zero indicates that these features are captured at time $t_0$. On the other hand, the vector $\gamma_k$ has a control nature and contains the on/off switching status of each DC. Assume $n_c$ DCs. Then, $\gamma_k=[q_1,...,q_{\ell},...,q_{n_c}]$, where $q_{\ell} \in \{0,1\}$ is the on/off switching status of the $\ell$-th DC. The subscript $k$ is the combination index. There are $2^{n_c}$ possible on/off switching combinations, thus, $k \in \{1,...,2^{n_c}\}$.
\subsection{Switching combination selection}
The approximator $\hat{S}_{\infty}$ outputs the estimated TA using $y_0$ and $\gamma_k$ as inputs. If oscillation modes are excited, the vector $y_0$ is formed using the online measurements at time $t_0$. Note that the particular operating condition and particular disturbance, that has excited the oscillation modes, are intrinsically captured through the vector $y_0$. With $y_0$ given, the optimal vector $\gamma^*$ must be found such that the estimated TA is minimized. In other words, given $y_0$, the search for the optimal on/off switching combination is done only in a subspace involved with the system current state. More formally, this problem is formulated as, 
\begin{equation}
	\begin{gathered}
		\label{DynamicCorDataBased}
		\mathbf{\gamma}^{*} = \arg~ \min _{\gamma_{k} \forall k \in \{1,...,2^{n_c}\}  } \hat{S}_{\infty}(\mathbf{y}_{0},\mathbf{\gamma}_{k})\\
	\end{gathered}	
\end{equation}
This is a binary optimization problem and its complexity depends on $n_c$. For a small $n_c$, minimum seeking can be achieved by sequentially determining the on or off status for each DC that leads to a smaller TA. This is done by directly evaluating the approximator with a fixed $y_0$ and fixed $\gamma_k$. The optimal solution can be also obtained by exhaustive evaluation of all possible combinations. If $n_c$ is large, integer optimization solvers are required. 
\subsection{Coordination structure}
\subsubsection{Decentralized control}
\begin{figure}[!h]
	\centering
    \includegraphics[width = 3.5 in]{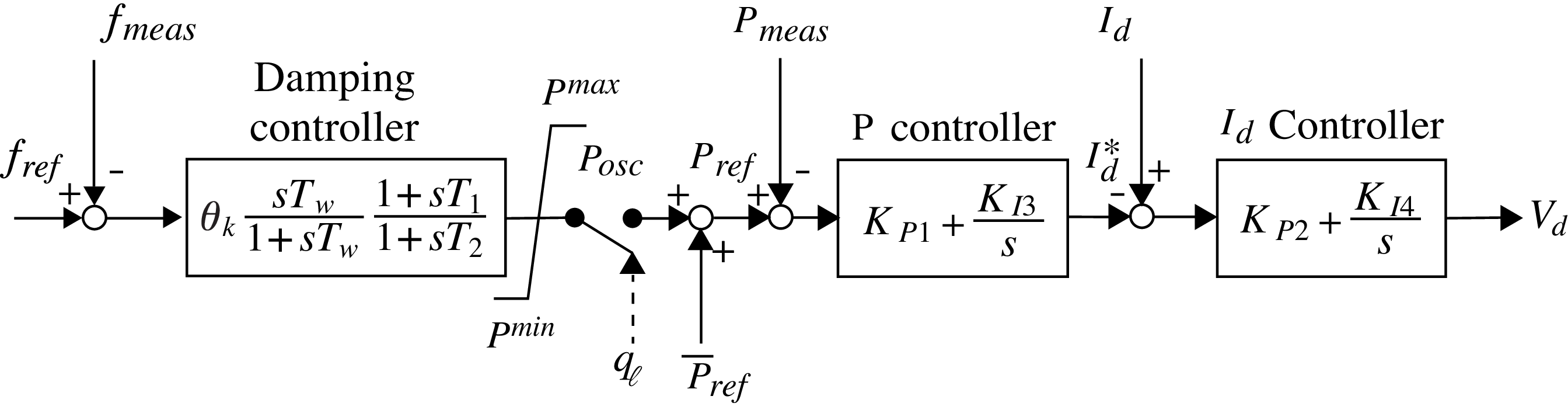}
	\vspace{-2ex}
	\caption{Active power control loop of $\ell$-th IBR.}
	\label{DampingCon}
\end{figure}

DCs are assumed to be implemented in IBRs, which are modeled as controlled voltage sources\cite{silva2020adaptive}. Assume that IBRs follow a quasi-stationary power reference, $\overline{P}_{ref}$, related to mid-term energy goals such as economic dispatch and load shifting. To provide damping control, the power output is modulated around the quasi-stationary reference targeting a specific oscillation mode through the supplementary reference signal $P_{osc}$. This signal originates from the corresponding damping control loop, which can be modeled as a traditional phase compensation damping controller with a damping constant, a washout filter, and a lead-lag compensator\cite{ZHANG201645}. The input signal and tuning parameters are selected based on control objectives and the type of modes, whether local for local modes or remote for wide-area damping control actuation against inter-area modes. For the $\ell$-th IBR, the total power reference is given by $P_{ref,\ell} = P_{osc,\ell} \cdot q_{\ell} + \overline{P}_{ref,\ell}$. If $q_{\ell}$ is set to zero, the DC is switched off, which leaves the IBR following $\overline{P}_{ref,\ell}$ only. Otherwise, the DC is turned on to provide damping control. This is shown in figure \ref{DampingCon}.
\subsubsection{Centralized control}
Once a fault has been cleared, a protection signal is transmitted to the centralized control at timestamp $t_{0}$ such that $\textbf{y}_{0}$ is formed from the measurements. The DNN is executed as described above to derive the optimal switching combination, $\gamma^*$, for this particular fault and operating condition. Subsequently, each switching signal, $q{\ell} \in \{0,1\}$, is transmitted by the centralized control to each DC, activating only those that are needed. If any of the IBRs participating in the coordination is out of service, the centralized control provides the best switching combination without taking into consideration the out-of-service elements.
\subsubsection{Supervisory System}
For a reliable implementation, the supervisory system will account for abnormal data conditions, i.e., PMU data loss, loss of input measurements, and data arrival inconsistencies due to time delays—PMU asynchronism. The input correction and data alignment are performed online but asynchronously with the coordination. This means that the data is constantly filtered by the supervisory mechanism before feeding into the coordination. When a signal is detected to be abnormal, the defective data correction method pre-process the measurements using historical and online data.
The authors further elaborated on this topic and proposed three different methods: a statistical-value-based method, which makes use of mean and median values of historical data; a minimal error-based method, which aims to derive the value with minimal prediction error from the known entries of the historical training input data; and a DNN-based adaptive method, which uses online and historical data for fast training and correction in real-time. For the sake of brevity, further details and the validation of these methods within this application can be found in \cite{Jingzi}.
\section{Setup and application}
\label{DDCimpl}
\subsection{Test system}
\begin{figure}[!b]
	\centering
    \includegraphics[width = 3 in]{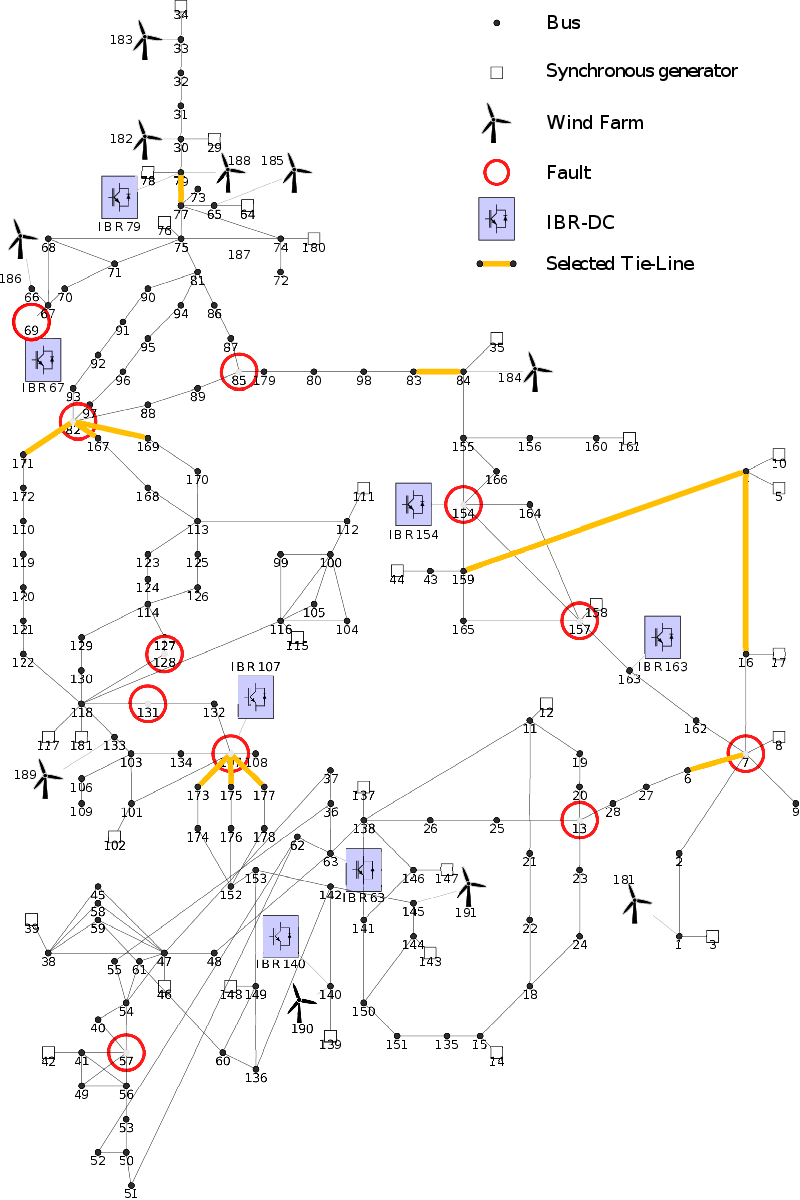}
	\caption{179-bus, 30-machine, 7-DC test system.}
	\label{wNAPS}
\end{figure}

A 179-bus, 30-machine, and 7-DC system is considered. This system is based on the Western North American Power System (wNAPS). A hypothetical case is assumed with 11 equivalent DFIG-based wind turbines installed in different buses throughout the system---see Fig. \ref{wNAPS}. For the different operating scenarios, the total instantaneous wind power generation falls in the range of 0-30\% of the total generation. Further details about the system and its data are found in references \cite{silva2020adaptive}, \cite{sun}. Although not resembling all actual conditions, the system model exhibits the five well-known inter-area modes of the wNAPS: ‘‘NS mode A’’, ‘‘NS mode B’’, ‘‘BC mode’’, ‘‘EW A’’ and ‘‘Montana mode’’, among others \cite{trudnowski2013pdci}. 
For the sake of analysis, the modes in this work have intentionally a more critically undamped behavior. 
\subsection{Feature vector}
One of the most important attributes of the proposed coordination is its adaptability to different operating conditions and faults---in what follows, the sub indices $i$ and $j$ are used to describe operating conditions and disturbances, respectively. To target each of these, the feature vector is further decomposed in two components as $\mathbf{y}_0=[\mathbf{y}_{01}, \mathbf{y}_{02}]^T$, Thus, the TA approximator becomes,
\begin{equation}
	\label{Mult_TA}
	\hat{S}_{\infty,i,j,k} = f_{_{DNN}}( \mathbf{y}_{0},\mathbf{\gamma}_{k}) = f_{_{DNN}}( \mathbf{y}_{01,i},\mathbf{y}_{02,j},\mathbf{\gamma}_{k})
\end{equation}

The first sub-vector, $\mathbf{y}{01}$, holds information highly related to the operating condition. Several variables can be considered; however, due to the fact that inter-area modes are the most dominant oscillations in the grid, the ideal variables must reflect the condition of the different electrical areas involved with these dynamics. For the wNAPS system, after conducting a power flow sensitivity analysis for different operating scenarios and taking into consideration its strong relation to power transfer among the main three electrical areas and the Canada-USA corridor in the north, transmission lines 6-7, 4-16, 4-159, 77-79, 83-84, 82-167, 82-169, 82-171, 107-173, 107-175, and 107-177 were selected. These lines are shown as thick solid lines in Fig. \ref{wNAPS}. The power flows in these lines are chosen as features, $\mathbf{p}{tl}$, and stored in $\mathbf{y}_{01}$.

The second sub-vector $\mathbf{y}_{02}$ contains information highly related to the location of the grid disturbances. Different disturbances can excite modes in different proportions. For example, in the case that more energetic modes are excited, with low damping and low frequency, the TA can have a value orders of magnitude higher than when less energetic modes are excited, with high damping and high frequency. Thereby, after a contingency, the frequency of the SG buses seems a suitable feature to capture the power grid response and the disturbance location. Note that the actual grid frequency is determined by the balance between demand and generation, which undergoes continuous variation. This makes pattern identification extremely challenging if the absolute frequency were to be used under different scenarios. To deal with this issue, the frequency deviation at SG buses with respect to a reference bus is used as feature, denoted as $\Delta \mathbf{\omega}_{sg}$, and stored in $\mathbf{y}_{02}$. This has shown to be appropriate, even with different pre-fault steady-state frequency values. 
\subsection{Data extraction}
DIgSILENT PowerFactory is used for dynamic simulations of all considered scenarios \cite{powerfactory2018powerfactory}. The PowerFactory/Python API, in Python 3.6, was used to run the time-domain simulations in the wNAPS under the set of specified operating conditions and disturbances as well as to automate the TA calculation and state collection. Algorithm 1 details a specific pseudocode to collect the entire data set. Consider a total of $n_o$ operating conditions, and $n_d$ disturbances/faults. Furthermore, define $t_{cl}$ as the fault clearing time, $t_a$ the activation time of the coordination, and $t_e$ the end time of the simulation. Thus, the entire collected data is stored in the super set $\Phi=\{\phi_1,...,\phi_s,...,\phi_{n_s}\}$, where $n_s=n_o \times n_d \times 2^{n_c}$, and $\phi_s=\{S_{\infty,i,j,k},\mathbf{y}_{01,i},\mathbf{y}_{02,j},\mathbf{\gamma}_{k}\}$ $\forall \ s \ \in \{1,...,n_s\}$. The collected data is captured in three batches. The first set $\mathbf{y}_{01,i}$ will capture the power flow through the selected transmission lines. This is taken in steady-state to provide information on the operating condition. Note if it were to be captured after a disturbance, this will be useless data as it will not resemble the same operating condition due to changes introduced by the disturbance. By keeping the element of the first set, a particular element of the second set $\mathbf{y}_{02,j}$ is formed. The time-domain simulation is invoked and temporarily paused at time $t_d$ to collect the frequency deviations. Later at time $t_a$, the simulation is temporarily paused once again to activate the switching combination of the DCs. At the end of the simulation, the subset $\phi_s$ is stored in the super set $\Phi$. 
\begin{algorithm}[h]
    \caption{Data collection}
    \label{Alg1}
   \scriptsize
    \begin{algorithmic}[1]
        \State Define $\mu_{i} \ \forall \ i  \in \{1,...,n_o\}$ \Comment{Op. conditions}
        \State Define $\upsilon_{j} \ \forall \ j  \in \{1,...,n_d\}$ \Comment{Disturbances}
        \State Define $\gamma_{k} \ \forall \ k  \in \{1,...,2^{n_c}\}$ \Comment{Switching comb.}
        \For {$each \  \mu_i$}
            \State Run power flow
            \State $\mathbf{y}_{01,i} = [\mathbf{p}_{tl}]$
            \For {$each \  \upsilon_{j}$}
                \For {$each \ \gamma_{k}$}
                    \State Run time-domain simulation
                    \If{$t=t_{cl}$}
                        \State $\mathbf{y}_{02,j} = [\Delta \mathbf{\omega}_{sg}]$
                    \EndIf
                    \If{$t=t_a$}
                        \State Activate coordination $\gamma_{k}$
                    \EndIf 
                    \If{$t = t_e$}
                        \State Calculate total action $S_{\infty, i,j,k}$
                        \State Store in $\Phi$ the subset $\phi_{s}=\{S_{\infty,i,j,k},\mathbf{y}_{01,i},\mathbf{y}_{02,j},\mathbf{\gamma}_{k}\}$
                    \EndIf
                \EndFor
            \EndFor
        \EndFor
    \end{algorithmic}
\end{algorithm}
\subsection{Training process}
For efficient and accurate learning, after the data extraction stage, the features $\mathbf{y}_{01}$ and $\mathbf{y}_{02}$ are standardized using a standard normal distribution (zero mean and unity standard deviation). To enhance performance and minimize biases, cross-validation is employed to determine the optimal hyperparameters for the DNN, including network structure, learning rate, number of epochs, and batch size. Then, the best set of parameters is chosen to train the DNN without selection bias and overfitting problems. This training process is done by applying stochastic gradient descent and backpropagation algorithm using the TensorFlow library in Python. 
\section{Case Study}\label{case_study}
The base case of the wNAPS considers no DCs but includes only conventional power system stabilizers in the SGs connected to buses 10, 29, 64, 69, 78, and 139. The adaptability of the data-informed coordination (DIC) to different disturbances is tested first and compared with the MBC. Later, the DIC is enhanced to include adaptability to operating conditions as well. Note that, for a given disturbance, the DIC determines those DCs that must be turned-on based only on system measurements ($\mathbf{y}_{01,i}$ and $\mathbf{y}_{02,j}$). At last, the time performance and time delay evaluation is presented.
\subsection{Coordination performance for different disturbances}
\begin{figure*}[b]
	\centering
    \includegraphics[width = 4.2 in]{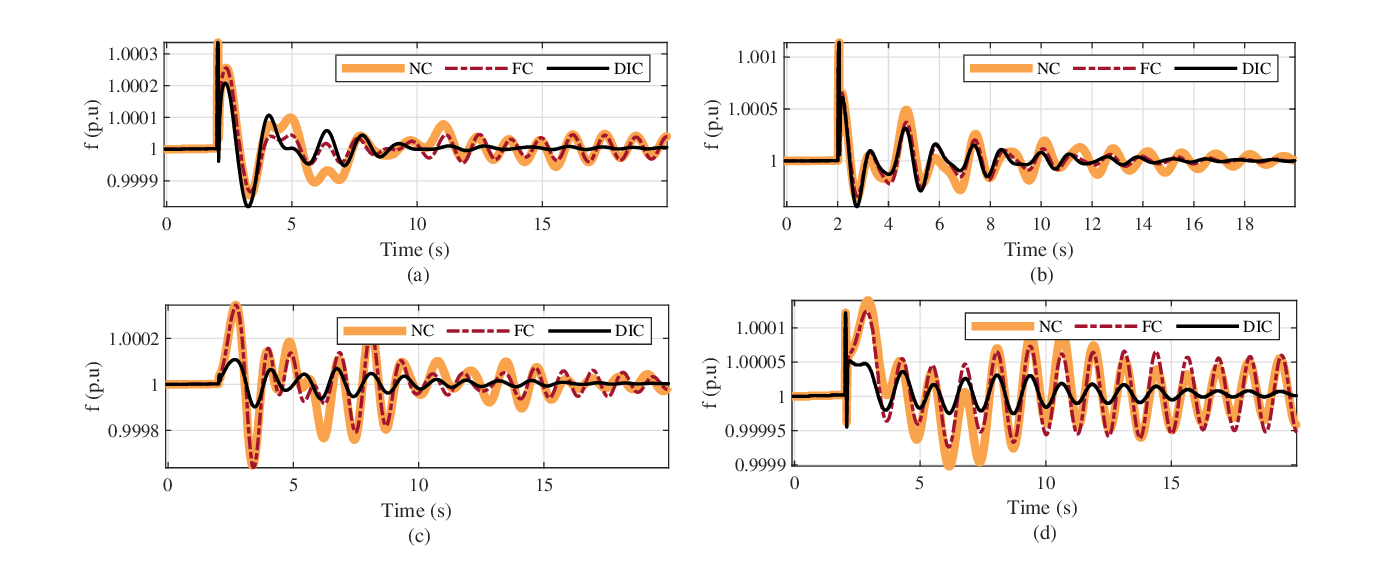}
	\vspace{-0.2 in}
	\caption{Frequency comparison for different control scenarios. Frequency at bus: (a) 117, (b) 14, (c) 161, and (d) 69.}
	\label{freqcom}
\end{figure*}

The DNN is trained considering 11 faults and 128 switching combinations ($n_{c}=7$ and $2^{n_{c}}=128$)---see Table \ref{DNN_table} for more details. Each nonlinear simulation considers a 3-phase fault to ground at t=2 s, which is cleared after 3 cycles. The fault locations were strategically distributed across the grid so that the dynamics of all major oscillatory modes were excited at some point. It is not necessary to consider all possible disturbances for the training process; this will be discussed later. These locations are depicted in Fig. \ref{wNAPS} using red circles. For simplicity, the DIC is activated at time $t_a=2.55$ s, which is 0.5 s after the fault is cleared. In practice, the DIC should be activated as soon as the optimal switching combination is obtained---no more than a few centiseconds with current computational capabilities. Fig. \ref{freqcom} shows the frequency at buses 117, 14, 161, and 69, for a fault on bus 57 with different control scenarios: No DCs activated (NC), orange curve; a fixed set of DCs activated (FC): DCs at buses 63, 79, 107, and 140 are turned on, maroon dashed curve; and the proposed DIC, black curve. Both DIC and FC improve the system damping, however, the former produces the best oscillation damping. 
The TA for these cases are $S_{\infty,DIC} = 0.1619$ and, $S_{\infty,FC} = 0.207$, which correspond to a $38.70\%$ and $21.60\%$ of TA reduction, respectively, compared to the base case.
\begin{table}[!htbp]
    \centering
    \caption{DNNs considerations}
    \label{DNN_table}
    \resizebox{1\textwidth}{!}{%
        \begin{tabular}{cccccc}
            \hline
            \textbf{DIC} & \textbf{No. of} & \textbf{No. of} & \textbf{Total No. of} & \textbf{Inputs} & \textbf{DNN} \\
            \textbf{Adaptability} & \textbf{Disturbances} & \textbf{Operating Conditions} & \textbf{Simulations} & \textbf{Measurements} & \textbf{Structure} \\
            \hline
            Disturbances & 11 & 1 & 1,408 & $\Delta\omega_{sg}$ & [37, 5000, 5000, 5000, 5000, 1] \\
            Operating Conditions and Disturbances & 11 & 13 & 18,304 & $\Delta\omega_{sg}$ and $\mathbf{p}_{tl}$ & [48, 3000, 3000, 3000, 5000, 5000, 5000, 1] \\
            \hline
        \end{tabular}%
    }
\end{table}

\begin{figure}[b!]
    \centering
    \begin{subfigure}{0.5\textwidth}
        \centering
        \hspace{-0.4in}
        \includegraphics[width=3in]{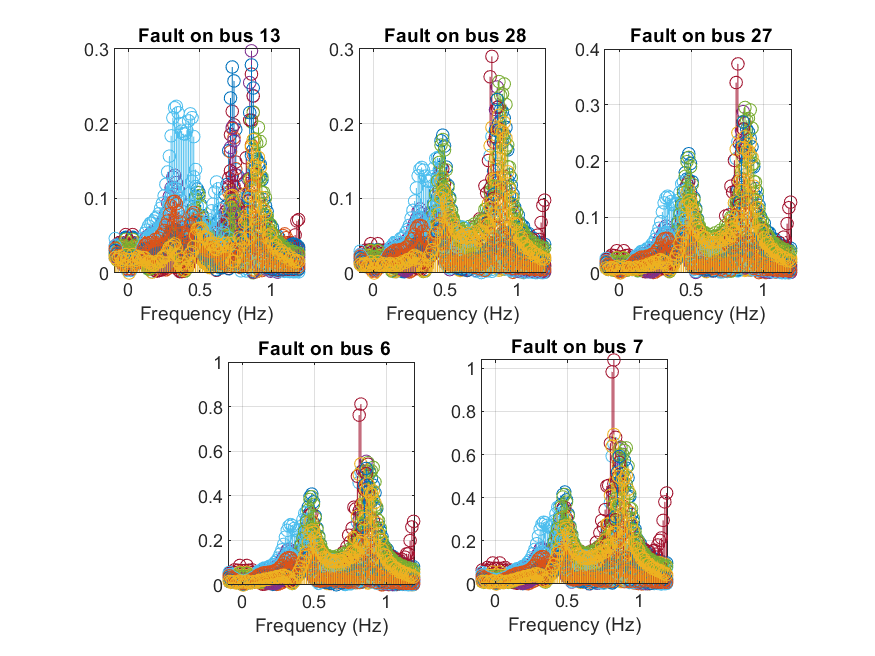}
        \vspace{0.1in}
        \caption{}
        \label{cora}
    \end{subfigure}%
    \begin{subfigure}{0.5\textwidth}
        \centering
        \includegraphics[width=2.8in]{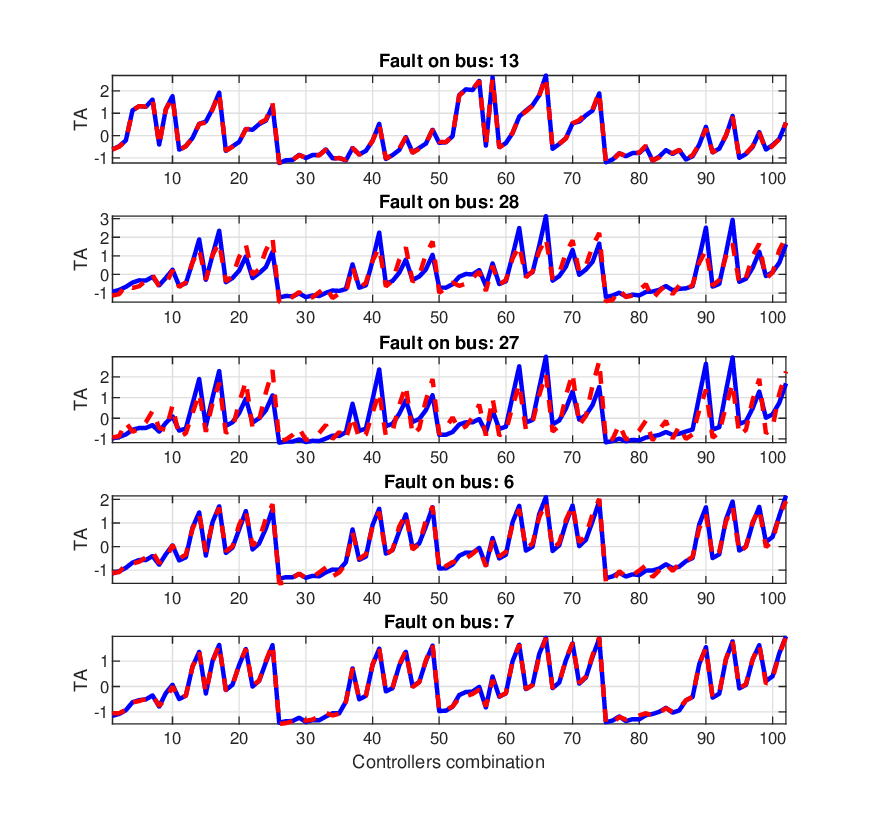}
        \caption{}
        \label{corb}
    \end{subfigure}
    \caption{Analysis of the corridor between buses 13-7: (a) Fourier spectrum, and (b) Total Action - actual vs. estimated (blue vs. red).}
    \label{Comp_MB_DB}
\end{figure}

As previously mentioned, the training process for this coordination does not necessarily need to consider a fault in every location within the grid. To illustrate this, the corridor between buses 13 and 7 is studied. When evaluating the dynamic response of the corridor, faults at buses 28, 27, and 6 exhibit similar behavior to those at the vertices. This is observed in Fig. \ref{cora}, where the Fourier spectrum of the system's response for the entire corridor is computed. The excited modes are the same for faults from bus 28 to 7, while at bus 13, the same modes remain excited; however, a new one appears around 0.70 Hz. This is important because, in the training process, only faults at buses 13 and 7 were considered.

When evaluating the total action, this behavior persists and can also be captured by the total action function approximator. This is shown in Fig. \ref{corb}, where the normalized actual TA is compared with the estimated one—blue curve vs. red curve, respectively. Each subplot corresponds to the evaluation of the total action for all controller combinations for one specific fault, where the x-axis represents the controller combination, and the y-axis represents the normalized TA value.
These results are further validated through the mean Spearman rank correlation coefficient, which, in this case, is 95.7\%. This index serves as a nonparametric measure of the strength and direction of the arbitrary monotonic association between two ranked variables: the actual and estimated TA. The findings demonstrate that, in certain locations, the grid exhibits similar dynamic behavior in response to different faults. Due to this behavior, strategic disturbances are used to generalize to new ones, providing acceptable estimations at fault locations that were not considered in the training set.
\subsection{Model-based vs data-informed adaptive coordination}
The proposed DIC has been compared with the MBC. Two short circuits are considered, one at bus 157, east side of the grid, and another at bus 69, north side. Fig. \ref{Comp_MB_DB} shows the oscillation energy for both cases, normalized with respect to the oscillation energy at time $t_0=2.05$ s. Table \ref{TotalComparisson} shows the comparison of the TA reduction. For both disturbances, DIC achieves the highest oscillation energy reduction, with a $73.53\%$ and $64.87\%$ for a fault at buses 157 and 69, respectively. For short-circuit at bus 69, FC even worsens the system dynamic response, leading to increased $E/E(t_0)$ and TA. It is a common practice to target a specific mode for PSSs or DCs tuning, however, some faults can excite modes very differently from the assumptions made during the tuning process. In those cases, a particular controller can be detrimental for the system stability. This new data-informed coordination is able to reduce the oscillation energy and TA, improving the damping of the system in all cases.
\begin{figure}[b!]
    \centering
    \begin{subfigure}{0.6\textwidth}
        \centering
        \includegraphics[width=\linewidth]{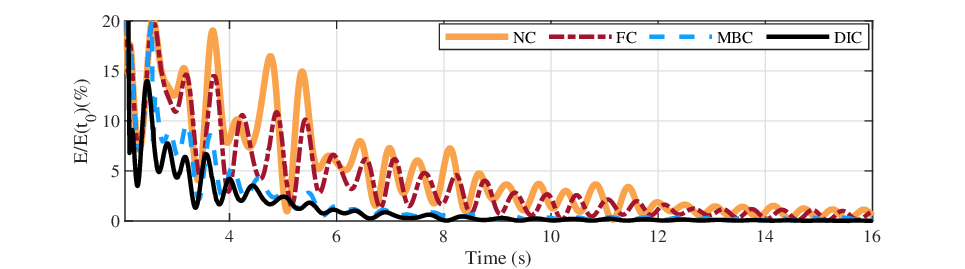}
        \caption{}
    \end{subfigure}
    \hspace{0.05\textwidth}
    \begin{subfigure}{0.6\textwidth}
        \centering
        \includegraphics[width=\linewidth]{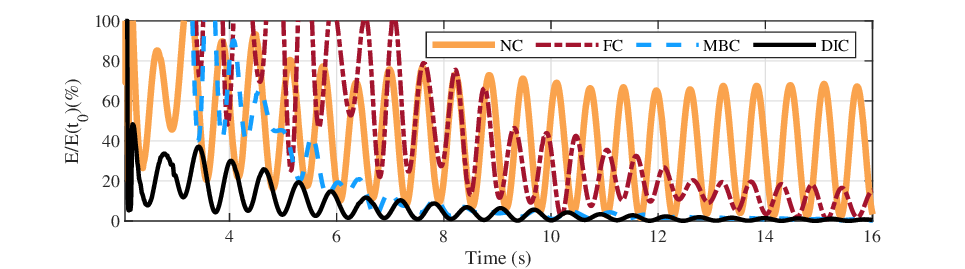}
        \caption{}
    \end{subfigure}
    \caption{MBC and DIC comparison for short-circuit at bus: (a) 157, and (b) 69.}
    \label{Comp_MB_DB}
\end{figure}
\begin{table}[!t]
    \caption{TA reduction per control scheme}
    \label{TotalComparisson}
    \centering
    \scriptsize
    \begin{tabular}{c|c|c|c|c}
        \hline
        \hline
        \multicolumn{1}{c}{} & \multicolumn{2}{c}{Short-circuit at bus 157} & \multicolumn{2}{c}{Short-circuit at bus 69} \\
        \hline
        \hline	  
        Coordination & Total Action & Reduction (\%) & Total Action & Reduction (\%) \\
        \hline
        No DC & 1.241 & - & 0.840 & -  \\
        \hline
        FC & 0.998 & 19.60 & 0.998 & -18.79  \\
        \hline
        DIC & \textbf{0.328} & \textbf{73.53} & \textbf{0.295} & \textbf{64.87}  \\
        \hline
        MBC & 0.500 & 59.72 & 0.473 & 43.65 \\
        \hline
    \end{tabular}
\end{table}
\subsection{Performance under different disturbances and operating conditions}
\begin{figure}[h!]
\centering
    \hspace{-0.37 in}
    \includegraphics[width = 3.1 in]{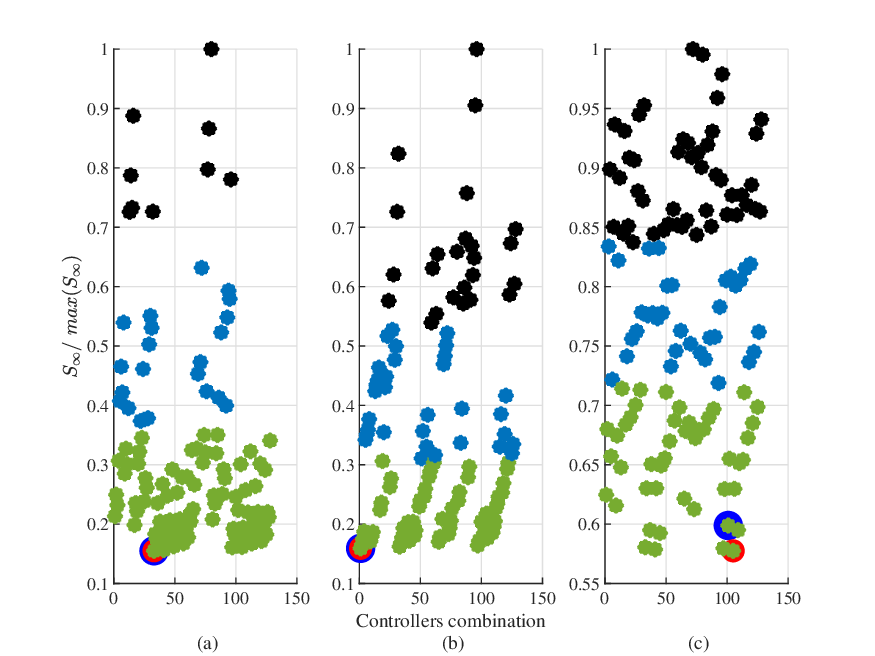}
	\caption{Controllers combination: a) Case-1, b) Case-2, and c) Case-3. }
	\label{comb_comp}
\end{figure}

Various demand levels, conventional generation, and wind power generation have been considered, leading to 13 distinct operating conditions for data extraction. As detailed in Table \ref{DNN_table}, a total of 18,304 simulations were conducted, which are all combinations of 11 disturbances, 13 operating conditions, and 128 controller combinations. The conditions for each individual simulation are the same as those used in the preceding section.
For the evaluation of controller performance, three cases are presented in this section: Case-1 considers operating condition 2, which is the base case, along with a fault at bus 107; Case-2 involves operating condition 10 with a fault at bus 154; and Case-3 deals with operating condition 6 and a fault at bus 85. Operating condition 10 exhibits a reduction of $10\%$ in demand, $9.6\%$ in thermal power generation, and $31\%$ in wind power generation compared to the base case. Operating condition 6 shows a reduction of $30\%$ in demand, $54\%$ in thermal generation, and no wind power generation at all.
\begin{figure}[b!]
\centering
    \includegraphics[width = 3.1 in]{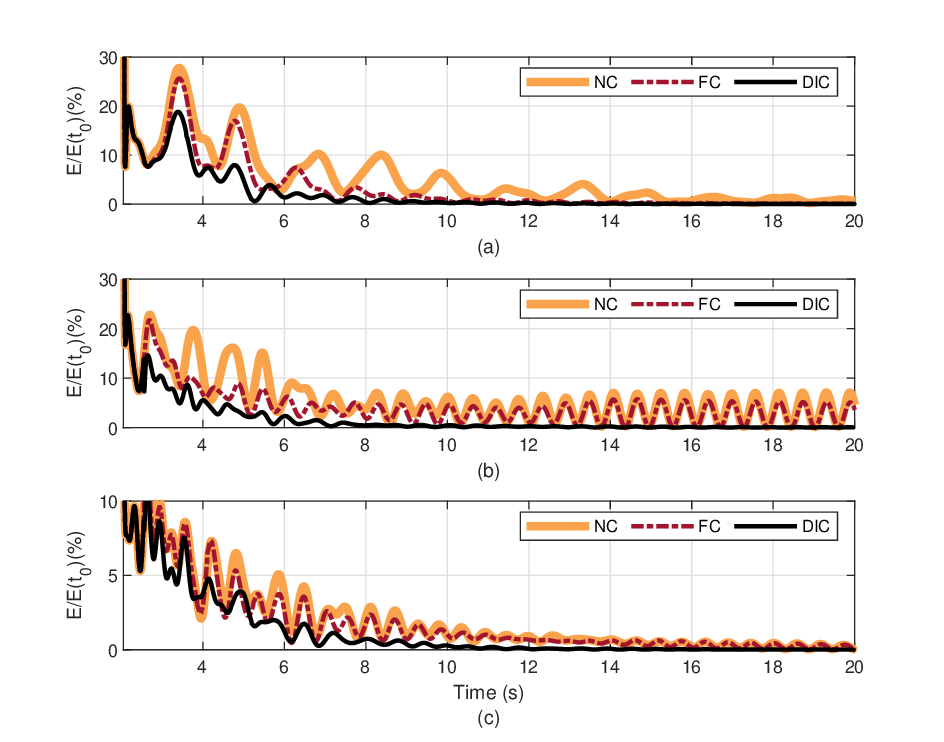}
	\caption{Oscillation energy comparison: a) Case-1, b) Case-2, and c) Case-3.}
	\label{OE_comp_ddc2}
\end{figure}
The proposed DIC exhibits the best performance among the control alternatives for all cases (see Fig. \ref{OE_comp_ddc2}). Even in Case-2, while FC struggles to stabilize the system, DIC completely annihilates the system oscillations. The controller is able to adapt to any disturbance and operating condition while keeping the DCs off in the initial state (pre-fault state). As indicated before, this is of tremendous importance, as the DIC commits the DCs only when they are required. To visualize the solution within the search space, Fig. \ref{comb_comp} highlights the activated coordination among all possible combinations for each case. In each case, there are 128 possible DCs combinations (x-axis) that produce different total action values $S_{\infty}$ (y-axis). Since the total action is a dimensionless quantity that depends on each particular case, this has been normalized over its maximum possible value for comparison. The red circle in each scatter plot corresponds to the coordination that provides the highest TA reduction, while the blue circle is the actual activated DCs coordination. In Case-1 and Case-2, the DIC is able to match the best coordination, while in Case-3 the combination activated is different than the best coordination, but remains one of the switching combinations that leads to lower TA.

\begin{figure}[t!]
\centering
     \hspace{-0.2 in}
    \includegraphics[width = 3.4 in]{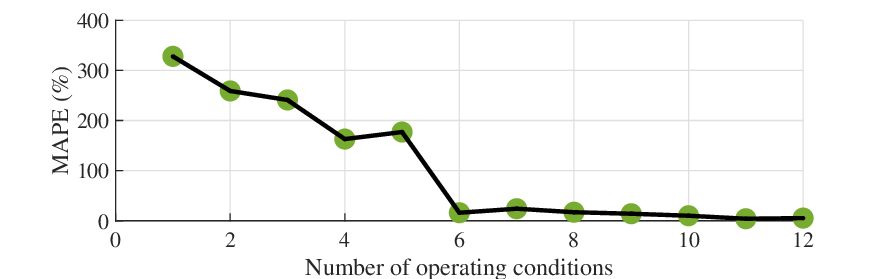}
	\caption{MAPE comparisson}
	\label{MAPE_com}
\end{figure}

To ensure effective coordination, it is crucial to generalize across diverse operating conditions and maintain a low estimation error in the TA magnitude, even for scenarios not covered during training. Following the law of large numbers, accuracy should naturally improve with an increasing number of operating conditions. This was evaluated by observing the mean absolute percentage error (MAPE) as the range of operating conditions was expanded in the training set. As shown in Figure \ref{MAPE_com}, the MAPE demonstrates noticeable improvement when incorporating more than six operating conditions, with a tendency to continue improving. This highlights the model's robustness in improving and adapting to other scenarios when a sufficient amount of operating conditions are considered.
\subsection{Time performance assessment}
A personal computer with intel(R) core(TM) i7-9750H CPU, NVIDIA GeFORCE GTX 1650 GPU, and 16 GB RAM has been used. The DNN and the search algorithm were constructed using TensorFlow 2.11.0 and Numpy 1.21.5, implemented on Python 3.9. Once the DNN has been trained, TensorFlow allows the prediction of a batch of inputs, in this case, the vector of input states $\mathbf{y}_{0}$ and the 128 controller's combinations. Then 128 predictions are provided, such that the combination with the smallest TA is selected. This process takes on average 78 ms per 1,000 tests. Moreover, contrary to the MBC, the DIC does not require previous knowledge of the current operating condition, this is, the current system conditions do not need to match with the information stored in the buffer of state matrices, eigenvectors, and their derivatives. 
\subsection{Time delay assessment}
\begin{figure}[h!]
\centering
     \hspace{-0.2 in}
    \includegraphics[width = 3.8 in]{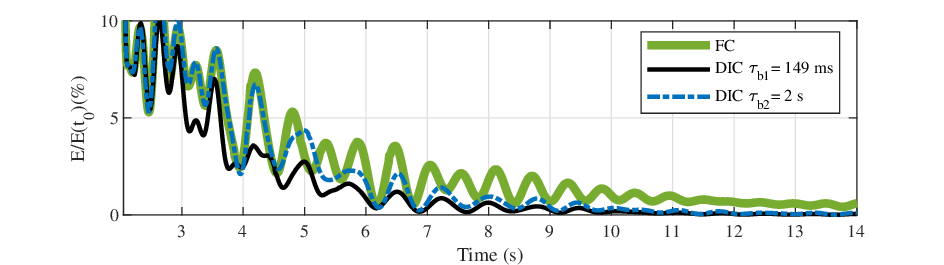}
	\caption{Time delay comparison.}
	\label{TimeDEl}
\end{figure}

The DIC is a discrete control that is in charge of optimizing the use of the continuous DCs by commissioning their status in the first few ms after a disturbance. Due to this, the evaluation of time delays can be done in a wider time frame. In this work, the parameter $\tau_{b}$ represents the combined effect of computation and communication delays. For an optimistic scenario, $\tau_{b1}$ is set to 149 ms. This value takes into consideration the time performance of the DIC and the reported delays associated with PMU communication and command handling, as reported in the Pacific DC Intertie wide-area damping controller \cite{pierre2019design}; and the pessimistic case, which is arbitrarily set to $\tau_{b2}=$ 2 s. Fig. \ref{TimeDEl} depicts the oscillation energy for the Case-3 discussed in the previous section, but now comparing the time delay scenarios. While the optimistic delay scenario yields the most significant reduction in oscillation energy, achieving a 41.4 \% decrease compared to the FC, it's worth noting that even the pessimistic case results in an enhanced dynamic response, with a total action reduction of 23.8 \%. Similar to the MBC, the DIC remains robust even in the worst-case delay scenario (communication loss), as it does not worsen the system's dynamic performance of the base case. Thus, the proposed adaptive coordination scheme is robust against time delays, improving satisfactorily the dynamic response of the system even in the worst considered scenarios. 
\section{Final remarks}
\label{disc}
While a DNN was considered as nonlinear function approximator, the concept is not limited only to this type of machine-learning model. In a realistic implementation, coordination should aim for generalization, utilizing a reduced but accurate set of inputs, maintaining simplicity in general maintenance, and providing accessible update capabilities. Physics-informed models, convolutions, or graph neural networks are alternatives that can capture complex input relations or extract features to reduce the number of inputs and improve the system's generalization \cite{spyros}. Alternatively, reinforcement learning can facilitate the retraining process and has shown to be a better solution for decision-making tasks \cite{chen2022reinforcement}. The core idea and framework behind coordination will be maintained; however, the AI model could be updated to a more robust solution.

Currently, the power grid industry is adopting guidelines from other industries for the deployment of AI-based solutions. In general, the deployment workflow considers different main steps \cite{paleyes2022challenges}. Each of these steps is a tailored process for specific problems and industries. For the power industry, some of the principal challenges include the quality and quantity of datasets, dependence on data extracted from time domain simulations, and the very restricted capability to validate solutions in a real grid before deployment \cite{alimi2020review}. These three topics remain open questions when considering the proposed coordination. To address this, future research will consider the use of hardware with larger computational capacity, defining a process to obtain a representative set of data that suffices for coordination generalization in a real grid, a more robust machine-learning algorithm, and implementing a testbed system to facilitate data extraction and test the coordination's behavior with hardware in the loop.
\section{Conclusion}
\label{conclusion}
This work proposes a novel data-informed adaptive coordination of damping controllers based on the total action performance index. The coordination framework uses wide-area measurements to adaptively decide the on/off status of several damping controllers enabled in IBRs. The decision-making process is done using a multivariate function approximation, allowing the selection of the controller combination that minimizes the total action to optimally increase the system damping. The results demonstrate the system's performance across different operating conditions, disturbances, and considerations for time delay. Unlike the model-based coordination procedure, this new framework provides accurate coordination without requiring any stored information or knowledge of the current operating condition. As a result, the proposed approach becomes faster, more accurate, and more dynamically adaptive. Additionally, it allows more appropriate use of IBRs such as PVs, WTs, and BESS, committing them to damping control actions only when required and thus avoiding curtailment/waste of energy. The data-driven philosophy of this approach makes it ideal for constantly adapting to operating conditions and disturbances. From the perspective of an operator, this is a non-intrusive control scheme since it does not require access to the tuning parameters of each damping controller. By only controlling the on/off status of well-designed offline IBR-DCs, the coordination can quickly adapt to the system's disturbance, dampen out the electromechanical oscillations, and improve the dynamic system response.  

\appendix


 \bibliographystyle{elsarticle-num} 
 \bibliography{DDC_SEGAN_V0}





\end{document}